\documentclass[preprintnumbers,amsmath,amssymb,prd,floatfix,twocolumn,superscriptaddress,nofootinbib]{revtex4}
\usepackage{graphicx}
\usepackage{epsfig}
\usepackage{bm}
\usepackage{amsfonts}

\begin{document}

\title{Thermodynamics of phantom energy in the presence of a Reissner-Nordstr\"{o}m
black hole}

\author{Mubasher Jamil}
\email{mjamil@camp.nust.edu.pk} \affiliation{Center for Advanced
Mathematics and Physics,\\ National University of Sciences and
Technology, H-12, Islamabad, Pakistan}

\author{Ibrar Hussain}
\email{ibrar.hussain@seecs.nust.edu.pk} \affiliation{School of
Electrical
Engineering and Computer Science,\\
National University of Sciences and Technology, H-12, Islamabad,
Pakistan}

\author{M. Umar Farooq}
\email{mfarooq@camp.nust.edu.pk} \affiliation{Center for Advanced
Mathematics and Physics,\\ National University of Sciences and
Technology, H-12, Islamabad, Pakistan}

\begin{abstract}
\centerline{\bf Abstract}In this paper,  we study the validity of
the generalized second law (GSL) in phantom dominated universe in
the presence of a Reissner-Nordstr\"{o}m (RN) black hole. Our study
is independent of the origin of the phantom like behavior of the
considered universe. We also discuss the GSL in the neighborhood of
transition from quintessence to phantom regime. We show that for a
constant equation of state parameter, the GSL may be satisfied
provided that the temperature is proportional to de Sitter
temperature. It is shown that in models with (only) a transition
from quintessence to phantom regime the generalized second law does
not hold in the transition epoch. Next we show that if the phantom
energy has a chemical potential, then the GSL will hold if the mass
of black hole is above from a critical value.
\newline
\newline
\textbf{Keywords:} Phantom energy; Reissner-Nordstr\"{o}m black
hole; Generalized second law of thermodynamics.
\end{abstract}

\author{}
\maketitle

\newpage

\section{Introduction}

The discovery that current era of the universe is in an accelerating
expansion phase obtained from many cosmological observations, such
as the type Ia supernova (SN Ia), Wilkinson Microwave Anisotropy
Probe (WMAP), the Sloan Digital Sky Survey (SDSS)
\cite{Riess1,Ben,Riess2,Eis} etc., is one of the most outstanding
achievements in modern cosmology. This unusual phenomenon provides
an intense interest to the scientific community to understand
(explore) the nature of the hidden force responsible for this
accelerating expansion. It is assumed that this acceleration is
mainly due to the presence of unusual stuff dubbed "dark energy",
which possesses positive energy density $\rho >0$ and negative
pressure $p<0$ which induce repulsive gravity \cite{Car,Zhang}.
Except for negative pressure, we do not know the other components
and properties of this mysterious form of energy. Despite the strong
observational evidence for the existence of dark energy, we have no
idea about how the dark energy evolves.

To explain the evolutionary behavior of the dark energy, various
models have been proposed. In all these models, the dark energy is
characterized by the equation of state $\omega =p/\rho $ ($p$ and
$\rho $ are the pressure and energy density of the dark energy
respectively) which facilitates us to understand the nature of dark
energy that accelerates the universe.

An unknown force (dark energy) which explains the accelerated
behavior of the universe is usually represented by a cosmological
constant which is nothing but a vacuum energy. However, to explain
the cosmic expansion, one requires the value of $\Lambda $ to be of
the order of $10^{-120},$ which can not be explained by current
particle physics. This is commonly known as the cosmological
constant problem. On the other hand, the first year WMAP data
together with the 2dF galaxy survey and the supernova Ia data favor
the phantom energy which has the equation of state $\omega <-1$ over
the cosmological constant and the quintessence. A candidate for
phantom dark energy is a scalar field with the wrong sign for
kinetic energy term \cite{Cald}.

In the study of the present accelerated expansion of the universe
driven by phantom energy, one may face a crucial situation in which
the phantom energy density and the scale factor blows up in a finite
time called Big Rip. This Big Rip may be avoided by introducing the
effect of gravitational back-reactions which may end the phantom
dominated regime \cite{Pwu}. So in this view, we can introduce
horizons for the accelerating universe and associate entropy and
temperature to them \cite{Pcw1,Pcw2,Poll,Izq,Mohseni,Nojiri,Davis}.
In this way one can make thermodynamical interpretation of a system
comprising of phantom dark energy in the form of perfect fluid and
the cosmological horizon.

Babichev et al. \cite{Babi}\ have shown that black holes lose mass
by accreting phantom energy and finally disappear completely. As a
result their areas will go down along with their entropies. By
keeping this picture into mind, we proceed to investigate whether
the GSL of thermodynamics holds in this scenario. For the sake of
interest one may expect that if the parameters assigned to the
universe are supposed to be the same as that of the ordinary
thermodynamics parameters related to the physical system, then the
thermodynamics laws must hold true by considering the global picture
(i.e. by considering universe as an object).

One can take the present universe as one thermodynamical entity.
Gibbons and Hawking \cite{Gibb} firstly investigated the
thermodynamical properties for the de Sitter spacetime, the event
horizon and the apparent horizon of the universe coincide and so
there is only one cosmological horizon. It was shown that the
cosmological horizon area can be interpreted as an entropy (measure
of some one ignorance about information of the regions behind it)
and thermal radiations coming from the cosmological horizon. The
thermodynamical study of the universe has been extended to the
quasi-de Sitter space in \cite{Poll,Frolo,Wang}. If the apparent
horizon and the event horizon of the universe are different, it has
been shown that first law and second law of thermodynamics hold true
for the apparent horizon. On the other hand these laws break down in
the case of event horizon \cite{Wang}.

In this paper, we explore the thermodynamical behavior of a universe
containing a RN black hole and phantom energy ($\omega<-1$). The
universe is considered as a closed thermodynamic system with a
boundary of future event horizon. Assuming the temperature of the
phantom energy and the cosmic horizon is the same, we check the
validity of GSL whether the total entropy of all components of the
system is an increasing function of time. The phantom energy
interacts with the black hole which leads the black hole mass to
decrease. This, henceforth, violates the ordinary second law of
black hole thermodynamics. We determined the condition under which
the GSL holds. We then compute the GSL near the time of phantom
transition and discuss its implications. Next we study GSL by taking
phantom energy having a chemical potential, which leads to a certain
critical value of black hole mass above which the GSL will hold.
Finally we conclude the paper.

\section{RN bBlack Hole in the Phantom
Dominated Friedmann-Lemaitre-Robertson-Walker Universe and GSL}

The line element of spatially flat
Friedmann-Lemaitre-Robertson-Walker (FLRW) metric with scale factor
$a(t)$ is given by
\begin{equation}\label{1}
ds^2=-dt^2+a^2(t)(dx^2+dy^2+dz^2).
\end{equation}
The Hubble parameter is defined by $H=\dot{a}/a$, where the over-dot
denotes derivative with respect to the comoving time $t$. Here we
assume the cosmology with the equation of state $p=\omega\rho$. For
$\omega<-1$ we have the phantom fluid and for the accelerated
expansion of the universe i.e. $\ddot{a}>0$, $ \omega<-1/3 $.

The future event horizon, $R_h$, is defined by
\begin{equation}\label{2}
R_h(t)=a(t)\int\limits_t^\infty\frac{d\tau}{a(\tau)}.
\end{equation}
Eq. (\ref{2}) corresponds to the distance that light will travel
from the present time till far in the future. In a phantom energy
dominated universe, the universe lasts for a finite time $t_*$. At
the Big Rip singularity i.e. at time $t=t_*$, $\infty$ must be
replaced with $t_*$ in the integration. Hence the future event
horizon is a finite distance.

For a system in quintessence i.e. $-1<\omega<-1/3$ ($\dot{H}<0$),
the future event horizon satisfy $\dot{R}_h\geq0$ and for a phantom
dominated universe i.e. $\omega<-1$ ($\dot{H}>0$), $\dot{R}_h\leq0$.
If the phantom ends to quintessence phase, one may have
$\dot{R}_h\geq0$ even in the phantom dominated regime.

An entropy for the future event horizon is given by
\begin{equation}\label{3}
S_h=\pi R_h^2.
\end{equation}
The total entropy of the universe, $S$ can be obtained as a sum of
the entropy inside the horizon, $S_{in}$, and $S_h$ \cite{GSL}
\begin{equation}\label{4}
S=S_{in}+S_{h}.
\end{equation}
Here it is assumed that the perfect fluid is in thermal equilibrium
with the future event horizon, as required for FLRW universe model.
When the future event horizon is de Sitter horizon i.e. when the
spacetime (1) is de Sitter, then the temperature can be considered
as $T={H}/{2\pi}$. When $R_h\neq H^{-1}$ i.e. for a non-de Sitter
spacetime, it is assumed that the future event horizon temperature
is proportional to the de Sitter temperature \cite{Pcw1,mohseni1}
\begin{equation}\label{5}
T=b\frac{H}{2\pi},
\end{equation}
where $b$ is a constant.

In the presence of dark energy and dark matter, the RN black hole is
introduced inside the future event horizon. Further, it is assumed that the
mass of the black hole, $M$, is small such that the FLRW model is unaltered.
With the use of $\rho=3H^2/8\pi$, this condition becomes
\begin{equation}\label{6}
MH\ll \frac{R_h^3 H^3}{2}.
\end{equation}
In terms of black hole entropy $S_{bh}$ and entropy of the perfect
fluid $ S_d $ the entropy inside the horizon may be given as
\begin{equation}\label{7}
S_{in}=S_{bh}+S_d
\end{equation}
The rate of change of mass of the RN black hole (due to phantom
energy accretion) is \cite{Babi}
\begin{eqnarray}\label{8}
\dot M&=&4\pi A_1M^2(\rho+p)\nonumber \\
&=&-A_1M^2\dot H.
\end{eqnarray}
where $r_h$ is the horizon of the black hole and $A_1$ is a constant. In
terms of Hubble parameter, $H$, the mass of the black hole may be given as
\begin{equation}\label{9}
M=\frac{1}{B+A_1H},
\end{equation}
where $B$ is a constant.

The black hole entropy is $S_b=4\pi M^2$, thereby
\begin{equation}\label{10}
\dot{S_{bh}}=-32\pi A_1M^3\dot{H}.
\end{equation}

The first law of black hole thermodynamics relates the entropy of the
phantom fluid to the energy and pressure
\begin{equation}\label{11}
TdS_d=dE+pdV=(\rho+p)dV+Vd\rho,
\end{equation}
where $V=4\pi R_h^3/3$ is the inside volume and $E=\rho V$ is the
total energy inside the future event horizon. Eq. (\ref{11}) yields
\begin{equation}\label{12}
T\dot {S}_d=\dot HR_h^2.
\end{equation}
To satisfy the GSL we must have
\begin{equation}\label{13}
\dot{S}_f+\dot{S}_{bh}+\dot{S}_h\geq0.
\end{equation}
This gives
\begin{eqnarray}\label{14}
&&\dot H\Big[\frac{R_h^2}{H}-2A_1bM^2\Big\{ M-2\sqrt{M^2-Q^2}-\frac{2M^2}{%
\sqrt{M^2-Q^2}} \Big\} \Big]\nonumber\\&&+ bR_h\dot{R}_h\geq0.
\end{eqnarray}
The above inequality is satisfied if the quantity inside the square
brackets is positive. In other words
\begin{equation}\label{15}
R_h^2\geq2A_1bHM^2\Big( M-2\sqrt{M^2-Q^2}-\frac{2M^2}{%
\sqrt{M^2-Q^2}} \Big).
\end{equation}
Note that in the above analysis, we assumed that the black hole is
non-extremal i.e. the black hole contains small electric charge
compared to the corresponding magnitude of mass.

\subsection{GSL Near Transition Time}

The astrophysical data favors the transition of the dark energy
state parameter from the sub-negative to super-negative values
around $-1$ \cite{Cald}. Following this observation, we check the
validity of the GSL near the transition time $t_0$ ($w=-1$) in the
presence of the RN black hole. In the phantom regime $\dot{H} > 0$
and in the quintessence regime we have $\dot{ H} < 0$, therefore if
the Hubble parameter has a Taylor series at transition time, which
is taken to be at $t_0 = 0$, $\dot{H}(0)=0$ and this is done by
applying the Taylor series expansion of the Hubble parameter about
$t_0=0$, we obtain
\begin{equation}\label{34}
H=h_0+h_1t^a,\ \ h_0=H_0,\ h_1=\frac{1}{a!}\frac{d^aH}{dt^a},
\end{equation}\
where $h_0 = H(t = 0)$ and $a$, a positive even integer number, is
the order of the first nonzero derivative of $H$ at $t = 0$. $h_1 =
H^{(a)}/a!$ and $H^{(a)} = d^aH/dt^a$. In the case of transition
from quintessence to phantom phase we must have $h_1 > 0$. Using
$\dot {R}_h=HR_h-1$, it can be shown that $R(t)$ has the following
expansions:
\begin{equation}\label{35}
R_h(t)=R_h(0)+(h_0R_h(0)-1)t+O(t^2),
\end{equation}
for $\dot{R}_h(0)\neq0$, and
\begin{equation}\label{36}
R_h(t)=R_h(0)\Big( 1+\frac{h_1}{a+1}t^{a+1} \Big)+O(t^{a+2}),
\end{equation}
for $\dot{R}_h(0)=0$, at $t=0$. Near the transition time, Eq.
(\ref{6}) reduces to $h_0^2R_h^3(0)\gg2M(0)$.

The condition of validity of GSL near the transition time, $t = 0,$
for $\dot{R}_h(0) = 0$, can be investigated by inserting $H = h_0 +
h_1t^a$ and (\ref{36}) into (\ref{14}):
\begin{eqnarray}\label{37}
&&a_1h_1\Big[ \frac{R_h(0)^2}{h_0}-2bA_1M(0)^2\Big( M(0)-2\sqrt{M(0)^2-Q^2}\nonumber\\&&-%
\frac{2M(0)^2}{\sqrt{M(0)^2-Q^2}} \Big) \Big]t^{a-1}+O(t^a)\geq0.
\end{eqnarray}
Note that $(a - 1)$ is an odd integer, therefore if the quantity in
the square bracket $\geq(\leq)0$, GSL is not respected in
quintessence (phantom) phase before (after) the transition. Indeed
the black hole mass $M(0)$, gives the possibility that GSL becomes
respected in the quintessence era before the transition.

\subsection{Phantom Energy With Chemical Potential and GSL}

We now proceed to study the GSL by assuming the dark energy having
chemical potential. Thermodynamical studies reveal several
interesting features of phantom energy: the temperature of the
phantom fluid without chemical potential is positive definite but
its co-moving entropy is negative. More recently, the thermodynamic
and statistical properties of phantom fluids were reexamined by
considering the existence of a non-zero chemical potential $\mu$. In
this case, it was found that the entropy condition, $S\geq0$,
implies that the possible values of $\omega$ are heavily dependent
on the value, as well as on the sign of the chemical potential
\cite{18}. In terms of the present day quantities (appearing below
with subscript 0), the energy density of a dark energy fluid can be
written as \cite{lima}
\begin{equation}\label{38}
\rho=\rho_0\Big( \frac{T}{T_0} \Big)^{\frac{1+w}{w}},
\end{equation}
whereas its entropy (including a chemical potential) reads
\cite{umar}
\begin{equation}\label{39}
S=\Big[ \frac{(1+w)\rho_0-\mu_0n_0}{T_0} \Big]\Big( \frac{T}{T_0} \Big)^{%
\frac{1}{w}}V,
\end{equation}
It thus follows that the total entropy of the system consisting of a
charged black hole plus a dark energy fluid reads
\begin{equation}\label{40}
S=\pi(M+\sqrt{M^2-Q^2})^2+\Big[ \frac{(1+w)\rho_0-\mu_0n_0}{T_0}
\Big]\Big( \frac{\rho}{\rho_0} \Big)^{\frac{1}{1+w}}V,
\end{equation}
where the first term represents the black hole entropy and the
second term is the phantom fluid entropy inside a co-moving volume
$V$, written in terms of the energy density. Now, due to the
accretion process, in an arbitrarily short time interval, the black
hole mass varies by $\Delta M$ and the phantom field energy varies
by $\Delta\rho$. Therefore, the total entropy variation within the
cavity takes the form
\begin{eqnarray}\label{41}
\Delta S&=&2\Delta M\Big[\pi(M+\sqrt{M^2-Q^2})\Big( 1+\frac{2M}{\sqrt{M^2-Q^2}} \Big)%
\Big]\nonumber\\&&+\frac{1}{1+w}\Big[ \frac{(1+w)\rho_0-\mu_0n_0}{T_0} \Big]\Big( \frac{%
\rho}{\rho_0} \Big)^{\frac{-w}{1+w}}V\Delta \rho.
\end{eqnarray}
For a phantom fluid modeled by a scalar field, only the kinetic term
contributes to the accretion, so that the energy conservation inside
the cavity implies \cite{cqg}
\begin{equation}\label{42}
\Delta M=-\frac{1}{2}(1+w)V\Delta\rho.
\end{equation}
Now, by inserting Eq. (\ref{42}) into Eq. (\ref{41}), we obtain an
expression for the total entropy variation of the black hole plus
dark energy
\begin{eqnarray}\label{43}
\Delta S&=&2\Delta M\Big[\pi(M+\sqrt{M^2-Q^2})\Big( 1+\frac{2M}{\sqrt{M^2-Q^2}} \Big)\nonumber\\&&-%
\frac{1}{(1+w)^2}\Big( \frac{(1+w)\rho_0-\mu_0n_0}{\rho_0T_0}
\Big)\Big( \frac{\rho}{\rho_0}
\Big)^{\frac{-w}{1+w}}\Big]\geq0.\nonumber
\end{eqnarray}
In the limit when the electric charge is negligibly small $M\gg Q$,
we have
\begin{eqnarray}\label{44}
M\Big( 6+\frac{Q^2}{2M^2} \Big)\geq M_{\text{crit}}&=&\frac{1}{\pi(1+w)^2}\Big( \frac{%
(1+w)\rho_0-\mu_0n_0}{\rho_0T_0} \Big)\nonumber\\&&\times\Big( \frac{\rho}{\rho_0} \Big)^{\frac{%
-w}{1+w}}.
\end{eqnarray}
Expression (\ref{44}) shows that GSL will hold if the black hole
mass is above a certain critical mass $M_{\text{crit}}$. In a
special case, if mass of the black hole is also very small and equal
in magnitude to the charge then the expression (\ref{44}), yields
\begin{equation}\label{45}
M\geq M_{\text{crit}}=\frac{2}{13\pi(1+w)^2}\Big( \frac{(1+w)\rho_0-\mu_0n_0}{%
\rho_0T_0} \Big)\Big( \frac{\rho}{\rho_0} \Big)^{\frac{-w}{1+w}}.
\end{equation}
Notice that the critical mass is negative when $w < -1$ and the
inevitable conclusion is that the process is physically forbidden.
In the point of view of \cite{papers}, such a result should be
physically expected since phantom fluids with negative temperature
cannot exist in nature. On the other hand, we see that for negative
values of $\mu_0$ there exists a positive critical mass above which
the black hole can accrete the phantom fluid. If we adopt the
results from Babichev et al. \cite{Babi}, where $\Delta M < 0$, the
condition to the mass is $M< M_{\text{crit}}$ . Therefore, only
black holes with mass below the critical mass can accrete phantom
energy.

\section{Conclusion}

In this work we have discussed the accretion of phantom fluids with
negative chemical potential by RN black hole. As we have seen, there
is a positive critical mass in order to enable the phantom
accretion. As physically expected, the GSL of thermodynamics
determines the thermodynamic viability of the whole process and the
amount of dark energy accretion. Phantom fluids with zero chemical
potential are not consistent because they require either a negative
entropy (which is microscopically unacceptable) or a negative
temperature (which needs a bounded spectrum which has not been
justified from any scalar field model).

\end{document}